\documentclass[prd,showpacs,nofootinbib,preprintnumbers,floatfix]{revtex4}  
\usepackage[plainpages=false, colorlinks=true, anchorcolor=blue, linkcolor=blue, citecolor=blue, bookmarks=false]{hyperref}
\usepackage{graphicx}  
\usepackage{dcolumn}   
\usepackage{bm}        
\usepackage{amssymb}   
\usepackage{natbib}
\usepackage{amsmath}   
\usepackage{longtable}  
\usepackage{float} \usepackage[utf8]{inputenc}
\usepackage{tasks}

\begin{document}

\newcommand{\apjl}{Astrophys. J. Lett.}
\newcommand{\apjs}{Astrophys. J. Suppl. Ser.}
\newcommand{\aap}{Astron. \& Astrophys.}
\newcommand{\rthis}[1]{\textcolor{black}{#1}}
\newcommand{\aj}{Astron. J.}
\newcommand{\pasp}{PASP}
\newcommand{\araa}{Ann. Rev. Astron. Astrophys. } 
\newcommand{\aapr}{Astronomy and Astrophysics Review}
\newcommand{\ssr}{Space Science Reviews}
\newcommand{\mnras}{Mon. Not. R. Astron. Soc.}
\newcommand{\apss} {Astrophys. and Space Science}
\newcommand{\jcap}{JCAP}
\newcommand{\na}{New Astronomy}
\newcommand{\pasj}{PASJ}
\newcommand{\pasa}{Pub. Astro. Soc. Aust.}
\newcommand{\physrep}{Physics Reports}

\title{Galaxy clusters, cosmic chronometers and the Einstein equivalence principle}

\author{I. E. C. R. Mendon\c{c}a$^{1}$}\email{ramalho.isaac@fisica.ufrn.br}

\author{Kamal Bora$^{2}$}\email{ph18resch11003@iith.ac.in}

\author{R. F. L. Holanda$^{1,3,4}$}\email{holandarfl@fisica.ufrn.br}

\author{Shantanu Desai$^{2}$}\email{shntn05@gmail.com}

\affiliation{$^1$Departamento de F\'{\i}sica, Universidade Federal do Rio Grande do Norte,Natal - Rio Grande do Norte, 59072-970, Brasil}

\affiliation{$^2$ Department of Physics, Indian Institute of Technology, Hyderabad, Kandi, Telangana-502284, India }

\affiliation{$^3$Departamento de F\'{\i}sica, Universidade Federal de Campina Grande, 58429-900, Campina Grande - PB, Brasil}

\affiliation{$^4$Departamento de F\'{\i}sica, Universidade Federal de Sergipe, 49100-000, Aracaju - SE, Brazil}

\begin{abstract}

The Einstein equivalence principle in the  electromagnetic sector can be violated in modifications of gravity theory  generated by a multiplicative coupling of a scalar field to the electromagnetic Lagrangian. In such  theories, deviations of the standard result  for the cosmic distance duality relation, and a variation of the fine structure constant are expected and  are unequivocally intertwined. In this paper, we search for these possible cosmological signatures by using galaxy cluster gas mass fraction measurements and  cosmic chronometers. No significant departure from  general relativity is found regardless  of our assumptions about cosmic curvature or  a possible depletion factor evolution in cluster measurements.

\end{abstract}
\pacs{98.80.-k, 95.35.+d, 98.80.Es}

\maketitle

\section{Introduction}

Despite  General Relativity (GR)  been corroborated in several Solar System tests  as well as in a strong gravitational field tests including gravitational wave observations \cite{will,Smoot,hole,Boran}, several other astronomical observations only can be explained if  new ingredients are added in the nature. Namely, the  dark matter (DM) and dark energy (DE) are necessary to explain the observations of galactic velocities in galaxy clusters, the rotational curve of spiral galaxies and the  discovery of the accelerated expansion of the universe  via observations of Supernovae type Ia (SNe Ia)~\cite{Huterer}. Nevertheless, the nature, origin and dynamics of such new kinds of matter is still unknown \cite{Cald,Wein}.  Through the last century, alternative gravity theories have been proposed in order to accommodate such observations, such as, massive gravity theories \cite{volkov,koba}, modified Newtonian dynamic (MOND) \cite{mond}, $f(R)$ and $f(T)$ theories \cite{fR}, braneworld models \cite{randall,pomarol,langlois,brax},
Bekenstein-Sandvik-Barrow-Magueijo(BSBM) theory \cite{Beke,Sand,Barrow,Beke2} of varying $\alpha$ among others.  Several of these theories break the Einstein equivalence principle (EEP), which leads to explicit modifications of some fundamental constants of nature. Then, it is also the role of observational cosmology to have mechanisms to test whether these theories actually satisfy  all the observational constraints. For example, Ref.~\cite{Mohapi}  considered a general tensor-scalar theory that allows to test the equivalence principle in the dark sector by introducing two different conformal couplings to standard matter and to dark matter. The analysis did not show any significant deviations from GR. By considering degenerate higher order scalar-tensor  theories,  Ref.~\cite{Cardone} also discussed which constraints can be put on the  parameters of these theories by using galaxy cluster data and gravitational lensing.

On the other hand, the authors of Refs.\cite{hees,hees2} proposed a robust mechanism to explore possible cosmological signatures of the modifications of gravity via the presence of a scalar field with a multiplicative coupling to the electromagnetic Lagrangian. It was shown that, in this context, variations of the fine structure constant, violations of the distance duality relation, departures of the standard evolution of the cosmic microwave background (CMB) temperature law are intimately and unequivocally linked.  In such a framework, the breaking of the EEP occurs by introducing an additional term into the action, coupling the usual matter fields $\Psi$ to a new scalar field $\phi$, which is motivated by a wide class of theories of gravity (scalar-tensor theories of gravity) \cite{string,string1,klein,axion,axion1,fine, fine1,fine2,fine3,chameleon,chameleon1,chameleon2,chameleon3,fRL,hees3,Martins}. Briefly, the explicit form of the couplings studied by \cite{hees,hees2} are of the type \begin{equation}
S_{m}=\sum_i \int d^4x\sqrt{-g}h_i(\phi)\mathcal{L}_i(g_{\mu\nu},\Psi_i)\,,\label{action1} \end{equation}
where $\mathcal{L}_i$ is the Lagrangians for different matter fields $\Psi_i$ and $h_i(\phi)$ represents a non-minimal couplings between $\phi$ and $\Psi_i$. When $h_i(\phi)=1$ we recover GR. Such a coupling affects  all the electromagnetic sector, which results in the  non-conservation of the photon number along geodesics,  changing the expression for the luminosity distance, $D_L (z)$ from the standard relation, where $z$ is the redshift, and violating the so-called cosmic distance-duality relation (CDDR)~\cite{Ellis}, $D_L= (1+z)^{2}D_A$, where $D_A$ is the angular diameter distance.  Moreover, a variation of the  evolution  of the CMB radiation is also  expected due to  the non-conservation of the photon number. All these changes are closely related to the time evolution of $h(\phi(t))$ (see next section).

In the last few years, several works have  proposed and implemented tests of  observational consequences from the action posited  in  Eq.~\ref{action1}, which explicitly breaks the EEP in the  electromagnetic sector  \cite{holandaprd,holandasaulo,holandasaulo2,Martins2,Martins3,Martins4,Martins5,rodrigoc,Holanda5,Holanda6,Martins6,Bora}. These studies  used angular diameter distances  of galaxy clusters obtained via their X-ray surface brightness + Sunyaev-Zel'dovich effect~\cite{SZ}  observations,  galaxy cluster gas mass fractions, strong gravitational lensing, SNe Ia samples  and the Cosmic Microwave Background (CMB) temperature in different redshifts, $T_{CMB}(z)$. Recently, Ref.\cite{levi} considered a number of well–studied $f(T)$ gravity models and devised various observational  predictions of the  corresponding EEP induced violation of the distance duality relation. 
The validity of EEP in $f(T)$ gravity was tested using current measurements of the variation of the fine–structure constant \cite{Murphy,King}. 
In general, no significant deviation from general relativity was found in these works, although their  results do not rule out with high confidence level the non-standard  models under question. 

In this paper, we search for deviations from  GR by considering  the class of models that explicitly breaks the EEP in the electromagnetic sector. The cosmological data used for this purpose consist of  40 gas mass fraction measurements in the redshift range $0.078 \leq z \leq 1.063$ \cite{mantz} and  31 cosmic chronometer $H(z)$ data  in the redshift range $0.07  \leq z \leq  1.965$~\cite{Li} in order to derive the angular diameter distance to the clusters. The search for signatures of the breaking of  EEP is performed by using a deformed cosmic distance duality relation, such as $D_L D_A^{-1}(1+z)^{-2}=\eta(z)$ and  $ \alpha(z)=\alpha_0  \varphi(z)$ (where $\alpha_0$ is the current
value of the fine-structure constant). For this class of theories $\eta(z)^2 \equiv \phi(z)$ and we consider $\eta(z)=1+\eta_0z$ in order to obtain tight limits on $\eta_0$. We explore both flat and non-flat universes. We also consider both a constant and redshift-dependent depletion factor,(i.e. the ratio by which the gas mass fraction of galaxy clusters is depleted with respect to the
universal mean of baryon fraction). As a basic result, $\eta_0 \approx 10^{-2}$ is obtained, independent of any assumptions on  the cosmic curvature as well as the depletion factor.
 
The outline of this paper is as follows. In Section~\ref{methodology}, we briefly explain the theoretical idea and methodology adopted in this paper. We discuss the data used in this work in Section~\ref{data}.  Section~\ref{sec:analysis} covers the analysis and results of this work. Our conclusions are discussed in Section~\ref{sec:conclusions}.

\section{Methodology}

\label{methodology}
\subsection{Theoretical framework}

{ Refs. \cite{hees,hees2} explored a class of modified gravity theories characterized by a universal non-minimal coupling between an extra scalar field $\Phi$ to gravity, where the standard matter Lagrangian $\mathcal{L}_i$ and the scalar field $\Phi$ are represented by the action:}
\begin{eqnarray}
S= \int  d^4x \sqrt{-g} \Big[ f_i(\Phi) \mathcal{L}_i (g_{\mu \nu}, \Psi_i) +  \frac{1}{2\kappa}\left(\Phi  R-\frac{\omega(\Phi)}{\Phi} (\partial_\sigma \Phi)^2-V(\Phi) \right)\Big], \label{action}
\end{eqnarray}
Here, $\kappa=8\pi G$, where $G$ is the gravitational constant, $V(\Phi)$ is the scalar-field potential, $R$ is the Ricci scalar for the metric $g_{\mu \nu}$ with determinant $g$, $f_i(\Phi)$, and $\omega(\Phi)$ are arbitrary functions of $\Phi$. The matter Lagrangian for the non-gravitational fields $\Psi_i$ is $\mathcal{L}_i$. For the electromagnetic radiation, $\mathcal{L}_{EM}(g_{\mu \nu}, \Psi_{EM})$, where $\Psi_{EM}=A^\mu$ stands for the 4-vector potential. {By  the extremization of the action (\ref{action}), one obtains the Einstein field equations \cite{hees}:
\begin{equation}
R_{\mu\nu}-{1\over 2}g_{\mu\nu}R=\kappa \frac{f_i(\Phi)}{\Phi}T^i_{\mu\nu}+\frac{1}{\Phi}[\nabla_\mu \nabla_\nu - g_{\mu\nu}\square]\Phi+ \frac{\omega(\Phi)}{\Phi^2}\bigg[\partial_\mu\Phi \partial_\nu\Phi-{1\over 2}g_{\mu\nu}(\partial_\alpha\Phi)^2\bigg]-g_{\mu\nu}\frac{V(\Phi)}{2\Phi}\,,
\end{equation}
with the stress-energy tensor being given by $T^i_{\mu\nu}=(-2/\sqrt{-g})\delta(\sqrt{-g}\mathcal{L}_i)/\delta g^{\mu\nu}$. As one may see, the cases $f_i(\Phi)\neq 1$ and/or $\Phi\neq 1$ will represent  the EEP breaking, while} the limit $\Phi\to 1$, $f_i(\Phi)\to 1$, $\omega(\Phi)=0$ and $V(\Phi)=0$ corresponds to the standard result. Moreover,  $\omega(\Phi)= constant$ and $f_i(\Phi)=1$ stands for the Brans-Dicke theory \cite{BD}, and the pressuron theory \cite{hees3}, the dilaton \cite{string,string1,Martins} also follow from that action. The main results in the case where the electromagnetic field is the only matter field present into the action (\ref{action}) are discussed below. 

 The Lagrangian in the vacuum is given by \cite{hees2}:
\begin{equation}
\mathcal{L}_{EM}(g_{\mu \nu}, A^\mu)=-\frac{1}{4}F^{\mu\nu}F_{\mu\nu}\,,
\end{equation}
where $F^{\mu\nu}=\partial^\mu A^\nu - \partial^\nu A^\mu$  and we will consider a coupling $f_i(\Phi)=f_{EM}(\Phi)$. The modified Maxwell equations can be obtained by variation of the action with respect to the 4-potential $A^\mu$: 
\begin{equation}
\nabla_\nu \left(f_{EM}(\Phi)F^{\mu\nu}\right)=0.
\end{equation}
Now, we expand the 4-potential as $A^\mu = \Re \left\{ \left(b^\mu + \epsilon c^\mu + O(\epsilon^2) \right) \exp^{i \theta / \epsilon} \right\}$ (standard procedure in general relativity \cite{minazzoli2,EMM}) and use the Lorentz gauge, which leads to the usual null-geodesic at leading order. The next order of the modified Maxwell equations is given by
\begin{eqnarray}
        k^\nu \nabla_\nu b &=&-\frac{1}{2}b\nabla_\nu k^\nu -\frac{1}{2}bk^\nu \partial_\nu \ln f_{EM}(\Phi) \label{amplitude}\\
        k^\nu \nabla_\nu h^\mu &=&\frac{1}{2}k^\mu h^\nu\partial_\nu \ln f_{EM}(\Phi)
\end{eqnarray}    
where $h^\mu$ is the polarisation vector,  $b$ is the amplitude of $b^\mu=b h^\mu$  and $k_\mu \equiv \partial_\mu \theta$. The photons number conservation law  is written as:
\begin{equation}
    \nabla_\nu \left(b^2k^\nu\right)=-b^2 k^\nu\partial_\nu \ln f_{EM}(\Phi).
\end{equation}
 {The wave vector in the flat FRW metric in spherical coordinate is $k^\mu=(k^0,k^r,0,0)=(-k_0,k_0/a(t),0,0)$} and it can be showed that the quantity $K=b(t,r) r a(t) \sqrt{f_{EM}(\Phi(t))}$ is constant along a geodesic. 

As it is known, the flux of energy comes from the $T^{0i}$ component of the energy momentum tensor, being given by
\begin{equation}\label{flux_tmp}
 F_0 =\left|a_0 b^2 k^0 k^r\right| =\frac{k_r^2b^2 }{a^2_0}=\frac{k_r^2 K^2}{r_0^2 a^4_0f_{EM}(\Phi_0)}=\frac{C}{r_0^2a_0^4f_{EM}(\Phi_0)},
\end{equation}
where $C$ is a constant. The emitted flux is:
\begin{equation}
F_e=\frac{C}{r_e^2a_e^4f_{EM}(\Phi_e)},
\end{equation}
here,  the index $e$ corresponds to the emitted signal. The angular integral of this defines the luminosity $L_e$:
\begin{equation}
L_e=\frac{4\pi C}{a_e^2f_{EM}(\Phi_e)}.
\end{equation}
Therefore, the equation for the distance of luminosity is:
\begin{equation}
D_L=\left(\frac{L_e}{4\pi F_0}\right)^{1/2}=\frac{a_0}{a_e}a_0 r_0 \sqrt{\frac{f_{EM}(\Phi_0)}{f_{EM}(\Phi_e)}}=c(1+z)\sqrt{\frac{f_{EM}(\Phi_0)}{f_{EM}(\Phi(z))}}\int_0^z\frac{dz}{H(z)}.\label{DL}
\end{equation}
As one may see, this expression for $D_L$ is slightly modified for a non-minimal coupling $f_{EM}$ between the electromagnetic Lagrangian and an extra scalar field.

Since  the angular diameter distance $D_A$ is a purely geometric quantity, its equation is the same as in usual electromagnetism
\begin{equation}
D_A(z)=\frac{c}{(1+z)} \int_{0}^{z} \frac{dz'}{H(z')}.\label{DA}
\end{equation}
By comparing with Eq.~\ref{DL} we have:
\begin{equation}
\frac{D_L(z)}{D_A(z)(1+z)^2}= \sqrt{\frac{f_{EM}(\Phi_0)}{f_{EM}(\Phi(z))}}\equiv\eta (z)\label{DLDA}.
\end{equation}
The parameter $\eta(z)$ is related to $f_{EM}(\Phi(z))$ for convenience, when $\eta(z)=1$, the above relation is also known as the cosmic  distance duality relation (CDDR). This relation plays an essential role in cosmological observations and  it has been tested in different cosmological context in recent years \cite{distance,distance1,distance2,distance3,distance4,distance5,kamal}. {In fact, as commented earlier, the kind of coupling explored by Refs. \cite{hees,hees2} also leads to a time-variation  of the fine structure constant, $\alpha=\alpha_0 \varphi(z)$,  as well as a modification of the CMB temperature evolution law, $T_{CMB}(z)=T_0(1+z)^{1-\tau}$, where  these  variations  intimately and unequivocally intertwined with  each other. As discussed in Ref. \cite{hees} (see their equations (12) and (34)), if a possible departure from the CDDR validity is quantified by a $\eta(z)$ term, the consequent deviation in the  CMB temperature evolution law and the time-evolution of the fine structure constant are described by:
\begin{equation}
\label{alpha}
\frac{\Delta \alpha}{\alpha}= \varphi(z) -1 = \eta^2 (z) -1
\end{equation}
and
\begin{equation}
T(z)=T_0(1+z)[0.88+0.12 \eta^2(z)]. \label{T}
\end{equation} }
In the follows, we discuss the consequences of a such coupling on  the galaxy cluster gas mass fraction measurements in the X-ray band.

\subsection{Consequences on Galaxy cluster gas mass fraction measurements}

Usually, the galaxy cluster  X-ray gas mass fraction measurements are  used to constrain cosmological parameters from an equation  that depends on the CDDR, such as \cite{Allen,Ettori,Allen2,mantz}:
\begin{equation}
\label{GasFrac}
f^{obs}_{X-ray}(z)=A(z)\gamma(z)K(z)(\Omega_b/\Omega_M)\left[\frac{D_L^* D_A^{*1/2}}{D_L D_A^{1/2}}\right].
\end{equation}
Here, $K(z)$ is the calibration constant equal to $0.96 \pm 0.12$, which accounts for any bias in the gas mass due to bulk motion and non-thermal pressure in the  cluster gas~\cite{Allen,Ettori,Allen2,mantz}, $A(z)$ represents the angular correction factor which is close to unity, and $D_A$ is the angular diameter distance to each cluster. $\gamma(z)$ is the depletion factor. The asterisk denotes the corresponding quantities in fiducial cosmology ($\Omega_m = 0.3$ and $h = 0.7$). However, if there is a departure of $\eta(z)=1$ ($D_L(1+z)^{-2}/D_A=\eta(z)$), this quantity would be affected  and it must be rewritten as (see Ref.\cite{gonc} for details):
\begin{eqnarray}
\label{GasFrac3}
f^{obs}_{X-ray}(z) &=& A(z)\gamma(z)K(z)(\Omega_b/\Omega_M) \left[\frac{D_A^{*3/2}}{\eta D_{A}^{3/2}}\right].
\end{eqnarray}
On the other hand,  the gas mass fraction measurements  extracted from X-ray data are also affected by a possible departure of $\varphi(z)=1$ (see Ref.\cite{holandasaulo2}), such as:
\begin{equation}
f_{X-ray} \propto  [\varphi(z)]^{-3/2}.
\end{equation}  
By considering the context of the  class of theories explored by \cite{hees,hees2}, where there is a breaking in the Einstein equivalence principle in electromagnetic sector and the relation $\varphi(z)^{1/2}=\eta(z)$ is verified, one obtains:
\begin{equation}
f_{X-ray} \propto  \eta^{-3}.
\end{equation}
Then,  the quantity $f^{obs}_{X-ray}$ may still be deviated from  its true value by a factor $\eta^{-3}$, which does not have a counterpart on the right side in the Eq.(\ref{GasFrac3}) \cite{holandasaulo2}. Then, this expression has to be modified to:
\begin{eqnarray}
\label{GasFrac4}
f^{obs}_{X-ray}(z) &=& A(z)\gamma(z)K(z)(\Omega_b/\Omega_M) \left[\frac{\eta^{2}D_A^{*3/2}}{D_A^{3/2}}\right].
\end{eqnarray}
Finally, if we have the angular diamater distance  at the  galaxy cluster redshifts, it is possible to constrain  the $\eta$ parameter by using:
\begin{eqnarray}
\label{eta}
\eta^{2} &=&  \left[\frac{f^{obs}_{X-ray}(z)D_A^{3/2}}{A(z)\gamma(z)K(z)(\Omega_b/\Omega_M)D_A^{*3/2}}\right].
\end{eqnarray}
If $\eta=1$, there is no breaking of EEP in electromagnetic sector. In this paper, the angular diameter distance for each galaxy cluster in the analyzed sample (see next section) is obtained by using $H(z)$ measurement data and we consider both flat and non-flat universes. For $\Omega_b$ and $\Omega_M$ we use the Planck's values, such as: 
$\Omega_b=0.0493 \pm 0.0039$ and $\Omega_M=0.3153 \pm 0.0073$~\cite{Planck18}.

\section{Data Sample}
\label{data}

\subsection{Gas Mass Fraction}
For galaxy cluster gas mass fraction measurements in the X-ray band we use a sample of 40 galaxy clusters  in the redshift range $0.078 \leq z \leq 1.063$~\cite{mantz}. The gas mass fraction of these structures were obtained in spherical shells at radii near $r_{2500}$\footnote{This radii is that one within which the mean cluster density is 2500 times the critical density of the Universe at the cluster's redshift.}, rather than  integrated at all radii ($< r_{2500}$) as in previous works. By using robust mass estimates for the target clusters via weak gravitational lensing  the bias in the mass measurements from X-ray data arising by assuming hydrostatic equilibrium was calibrated~\cite{App}. An important factor when one uses this kind of measurements is the depletion factor, which is the ratio by which the gas mass fraction of galaxy clusters is depleted with respect to the
universal mean of baryon fraction. Here we consider two cases: a constant depletion factor and an evolving one such as  $\gamma(z) = \gamma_0(1+\gamma_1 z)$.

\subsection{Cosmic Chronometers}
\begin{figure*}[t]
    \centering
    \includegraphics[scale=0.4]{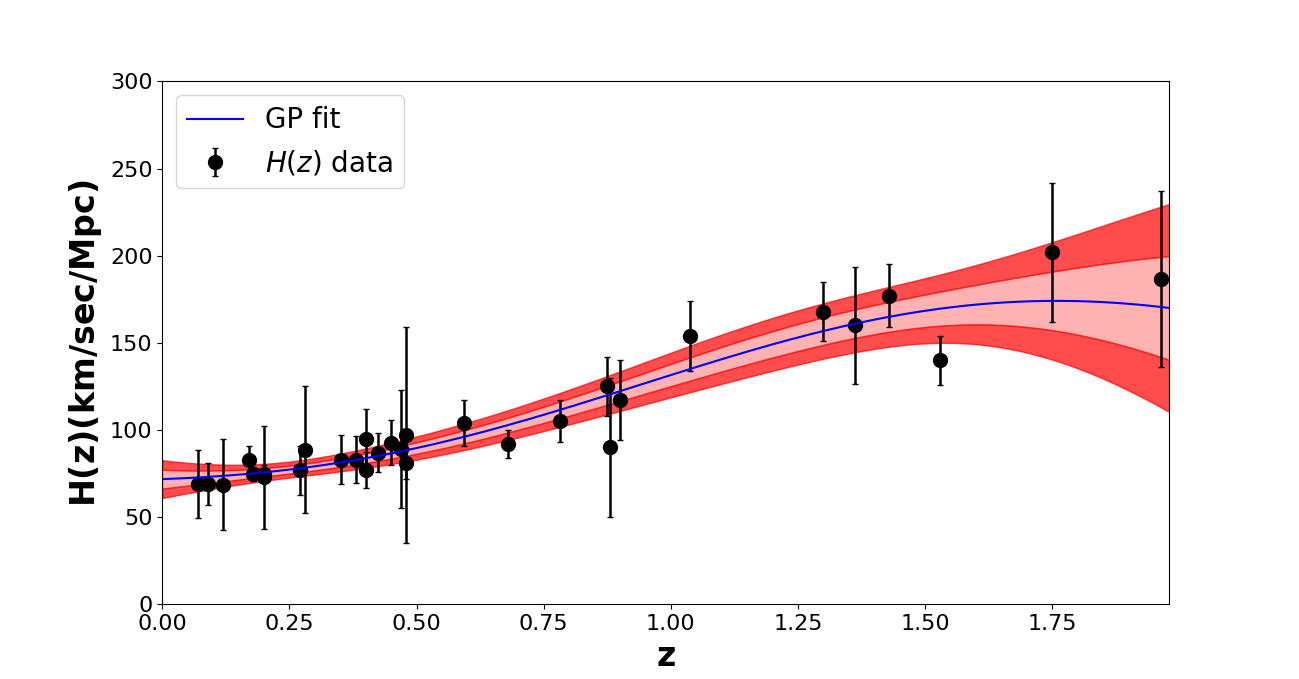} 
    \caption{Reconstruction of Hubble parameter at different redshift using Gaussian Processes. We used 31 $H(z)$ cosmic chronometers measurements compiled by \cite{Li}. The best GP fit is shown by a blue line along with $1\sigma$ and $2\sigma$ error bands.}
    \label{fig:fig2}
\end{figure*}

We use 31 cosmic chronometers $H(z)$ data from Ref.\cite{Li} in the redshift range $0.07 \leqslant z \leqslant 1.965 $ in order to derive the angular diameter distance to galaxy clusters (see Fig~\ref{fig:fig2}). As it is largely known, Cosmic Chronometers (CC) are one of the most widely used probes in Cosmology. Briefly,  if passively evolving galaxies at different redshifts are considered,  the age difference of the galaxies can be computed yielding in the Hubble parameter based on the following equation \cite{Sing,Jimenez}:
\begin{equation}
    H(z) = - \frac{1}{1+z} \frac{dz}{dt}.
\label{eq:chrono}
\end{equation}
The derivative term $dz/dt$  in Eq. (\ref{eq:chrono}) is obtained with respect to the cosmic time. The only assumption for the CC is the stellar population model, no cosmological model is considered. We also plot in Fig.(1) the reconstruction of Hubble parameter at different redshift using Gaussian Processes. This will be important to compute the comoving distance in galaxy cluster's redshifts of the previous sample.

As commented earlier, the main goal of this work is to obtain limits on $\eta_0$ parameter from the  $\eta(z)=1+\eta_0z$ function, which gives us information if there is some breaking in EEP in electromagnetic sector.  For this purpose, we use three different cases which we will discuss one by one in next section.


\section{Models Tested}
\label{models}
In all cases, we choose Gaussian Processes Regression~\cite{seikel} to reconstruct the comoving distance, $D_C$, at each galaxy cluster's redshift via $H(z)$ data (for more details, see~\cite{Sing,boraepjc,kamal}), where $D_C$ is:

\begin{equation}
D_C=c\int_{0}^{z}\frac{dz^{'}}{H(z')}.
\end{equation}

\subsection{A flat Universe with a constant gas depletion factor}
\label{case1}
In this case, the angular diameter distance for each galaxy cluster can be obtained from $D_C$ by:

\begin{equation}
D_{A} (z) =  \frac{D_C}{(1+z)},
\label{eq:daz}
\end{equation}
and the depletion factor is simply $\gamma(z)=\gamma_0$. Then, from Eq.(\ref{eta}), the limits on $\eta_0$ and $\gamma_0$ parameters can be done by using:

\begin{widetext}
\begin{equation}
    \label{eq:case1}
    \gamma_0(1+\eta_0z)^{2} =   \left[\frac{f_{gas}(z)}{K(z)A(z)}\right]     \left[\frac{\Omega_m}{\Omega_b}\right] \left(\frac{D_A}{D_A^*}\right)^{3/2} 
\end{equation}    
\end{widetext}

\subsection{A flat Universe with varying gas depletion factor}
\label{case2}

In this case, the angular diameter distance for each galaxy cluster is still obtained via Eq.(\ref{eq:daz}), but the depletion factor  is given by:

\begin{equation}
\label{gamma_parametric}
\gamma(z) = \gamma_0(1+\gamma_1 z).
\end{equation}
Then,  the limits on $\eta_0$, $\gamma_0$ and $\gamma_1$ parameters can be done by using:

\begin{widetext}
\begin{equation}
    \label{eq:case2}
    \gamma_0(1+\gamma_1z)(1+\eta_0z)^{2} =   \left[\frac{f_{gas}(z)}{K(z)A(z)}\right]     \left[\frac{\Omega_m}{\Omega_b}\right] \left(\frac{D_A}{D_A^*}\right)^{3/2} 
\end{equation}    
\end{widetext}

\subsection{A non-flat Universe with a constant gas depletion factor}
\label{case3}

Finally, in the last case we add the curvature parameter in analyses as a free parameter.  This is motivated by recent observations which hint at non-vanishing curvature~\cite{Divalentino}.
The angular diameter distance is related to the  comoving distance $D_C$ by following equations (for more details, see \cite{hogg99}),

\begin{widetext}
\begin{equation}
\label{cov_dist}
D_A(1+z)=
\begin{cases} 
D_H \frac{1}{\sqrt{\Omega_k}} \text{sinh}\left[\frac{\sqrt{\Omega_k} D_C}{D_H}\right] \hspace{1.3cm}  \Omega_k > 0 \\
D_C  \hspace{4.17cm}   \Omega_k = 0 \\
D_H \frac{1}{\sqrt{|\Omega_k|}} \text{sin}\left[\frac{\sqrt{|\Omega_k|} D_C}{D_H}\right]  \hspace{1cm}  \Omega_k < 0
\end{cases}
\end{equation}
\end{widetext}
where $D_H=c/H_0$. Then,  the limits on $\eta_0$, $\gamma_0$ and $\Omega_k$ parameters can be done by using:

\begin{widetext}
\begin{equation}
    \label{eq:case3}
    \gamma_0(1+\eta_0z)^{2} D_A^{-3/2} =   \left[\frac{f_{gas}(z)}{K(z)A(z)}\right]     \left[\frac{\Omega_m(z)}{\Omega_b(z)}\right] (D_A^*)^{-3/2} 
\end{equation}    
\end{widetext}

\begin{table*}[t]
\caption{\label{tab:table1}. Constraints on the parameter $\eta_0$ as discussed in Sec~\ref{models} along with the recent studies.}
    \centering
    \begin{tabular}{|l|c|c|c|r|} \hline
    \textbf{Dataset Used} & \boldmath$\eta_0$ & \textbf{Reference}\\ \hline 
        ADD + SNeIa  & $0.069 \pm 0.106$ & \cite{holandaprd}\\
        ADD + SNeIa + $T_{CMB}$ & $-0.005 \pm 0.025$ &\cite{holandasaulo}\\
         Gas Mass Fraction+SNeIa+$T_{CMB}$ & $-0.020 \pm 0.027$ & \cite{holandasaulo2}\\
         ADD+Gas Mass Fraction+SNeIa+$T_{CMB}$&$-0.012 \pm 0.022$ &\cite{rodrigoc}\\
          Galaxy clusters + SNeIa + $H(z)$ & $0.009\pm0.05$ & \cite{kamal}\\
               Gas Mass Fraction + Cosmic Chronometers \textbf{(Case I)} &  $\mathbf{-0.017_{-0.075}^{+0.077}}$ & \textbf{This work}\\
       Gas Mass Fraction + Cosmic Chronometers \textbf{(Case II)} & $\mathbf{-0.115_{-0.211}^{+0.362}}$ & \textbf{This work}\\
       Gas Mass Fraction + Cosmic Chronometers \textbf{(Case III)} & $\mathbf{0.081^{+0.389}_{-0.359}}$ & \textbf{This work}\\

      \hline 
      
    \end{tabular}

\end{table*}

\section{Analysis and Results} 
\label{sec:analysis}

\subsection{Constraints on $\eta_0$ parameter}
\textbf{Case A:}
In order to constrain the parameters $\eta_0$ and $\gamma_0$ of Eq.(\ref{eq:case1}), we maximize the likelihood equation given as below,

\begin{widetext}
\begin{equation}
    \label{eq:logL1}
   -2\ln\mathcal{L} =  \sum_{i=1}^{n}\frac{\left( \gamma_0(1+\eta_0z)^{2} - \left[\frac{f_{gas}(z)}{K(z)A(z)}\right]     \left[\frac{\Omega_m(z)}{\Omega_b(z)}\right] \left(\frac{D_A}{D_A^*}\right)^{3/2}\right)^2}{\sigma_{i}^2} + \sum_{i=1}^{n} \ln 2\pi{\sigma_{i}^2}. 
\end{equation}    
\end{widetext}
Here, $\sigma_i^2$ denotes the observational errors which is obtained by propagating the errors in $f_{gas}(z)$, $K(z)$, $\Omega_{m}$, $\Omega_{b}$, $D_A$, and $D_A^*$.

\textbf{Case B :}

Similarly, the constraints on the $\eta_0$ parameter along with $\gamma_0$ and $\gamma_1$ present in Eq.(\ref{eq:case2}) can be obtained by maximizing the likelihood distribution function, ${\cal{L}}$  given by

\begin{widetext}
\begin{equation}
    \label{eq:logL2}
   -2\ln\mathcal{L} =  \sum_{i=1}^{n}\frac{\left( \gamma_0(1+\gamma_1z)(1+\eta_0z)^{2} - \left[\frac{f_{gas}(z)}{K(z)A(z)}\right]     \left[\frac{\Omega_m(z)}{\Omega_b(z)}\right] \left(\frac{D_A}{D_A^*}\right)^{3/2}\right)^2}{\sigma_{i}^2} + \sum_{i=1}^{n} \ln 2\pi{\sigma_{i}^2}. 
\end{equation}    
\end{widetext} 
Here, $\sigma_i^2$ includes the error in $f_{gas}(z)$, $K(z)$, $\Omega_{m}$, $\Omega_{b}$, $D_A$, and $D_A^*$ which is calculated by error propagation of these quantities.

\textbf{Case C :}

We can put limits on $\eta_0$ parameter assuming $\Omega_k$ and $\gamma_0$ as the free parameters(see Eq.(\ref{eq:case3})). In this case, the likelihood function is given by,

\begin{widetext}
\begin{equation}
    \label{eq:logL3}
   -2\ln\mathcal{L} =  \sum_{i=1}^{n}\frac{\left( \gamma_0(1+\eta_0z)^{2} D_A^{-1.5} - \left[\frac{f_{gas}(z)}{K(z)A(z)}\right]     \left[\frac{\Omega_m(z)}{\Omega_b(z)}\right] \left({D_A^*}\right)^{-3/2}\right)^2}{\sigma_{i}^2} + \sum_{i=1}^{n} \ln 2\pi{\sigma_{i}^2}. 
\end{equation}    
\end{widetext} 
Here also, $\sigma_i^2$ represents the observational errors in $f_{gas}(z)$, $K(z)$, $\Omega_{m}$, $\Omega_{b}$, and $D_A^*$.

\subsection{Results}

\begin{figure*}[t]
    \centering
    \includegraphics[width=10cm,height=8cm]{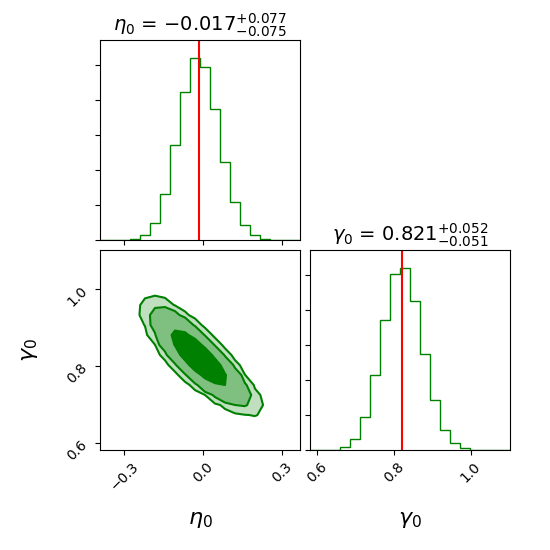}
    \caption{Constraints on $\eta_0$ and $\gamma_0$. The diagonal plots represent 1-D marginalized likelihood distributions of each parameter present in Eq.(\ref{eq:case1}) and off-diagonal contours are 68\%, 95\%, and 99\% 2-D marginalized confidence regions. This plot is generated via {\tt Corner} python module~\cite{corner}.}
    \label{fig:fig3}
\end{figure*}

\begin{figure*}[t]
    \centering
    \includegraphics[width=10cm,height=8cm]{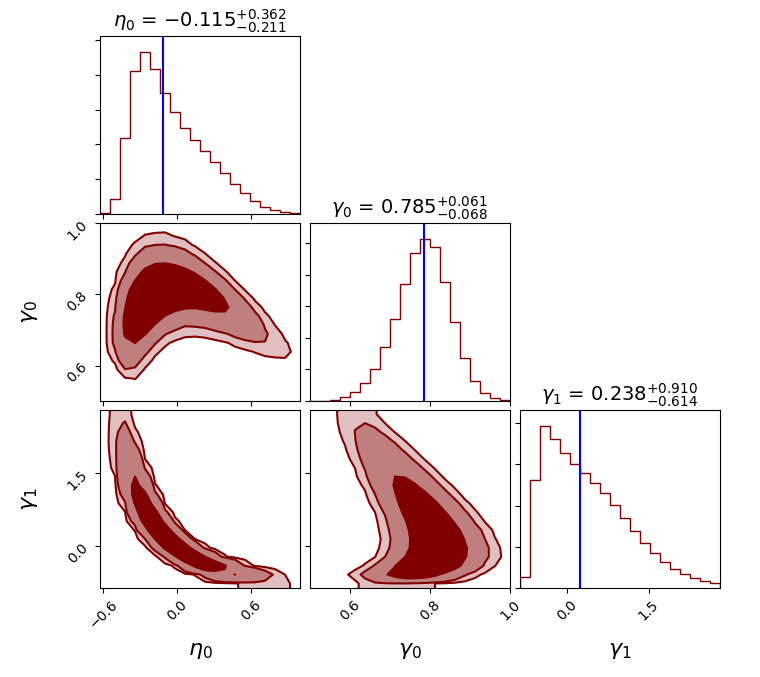}
     \caption{Constraints on $\eta_0$, $\gamma_0$, and $\gamma_1$. The diagonal plots represent 1-D marginalized likelihood distributions of each parameter present in Eq.(\ref{eq:case2}) and off-diagonal contours are 68\%, 95\%, and 99\% 2-D marginalized confidence regions.}
    \label{fig:fig4}
\end{figure*}

\begin{figure*}[t]
    \centering
    \includegraphics[width=10cm,height=8cm]{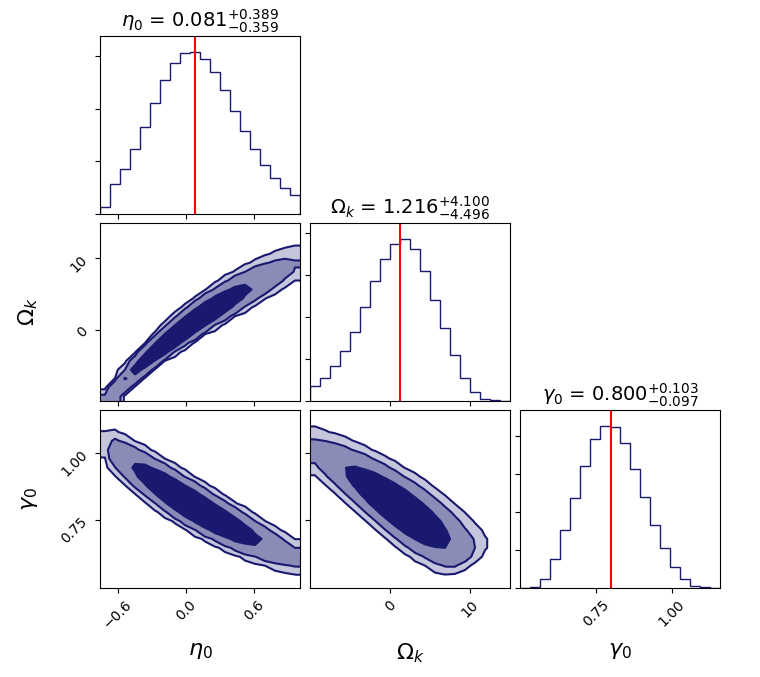}
    \caption{Constraints on $\eta_0$,  $\Omega_k$, and $\gamma_0$. The diagonal plots represent 1-D marginalized likelihood distributions of each parameter present in Eq.(\ref{eq:case3}) and off-diagonal contours are 68\%, 95\%, and 99\% 2-D marginalized confidence regions.}
    \label{fig:fig5}
\end{figure*}

To maximize the likelihood distribution functions given in Eq.(\ref{eq:case1}),(\ref{eq:case2}), and (\ref{eq:case3}), we use the $\tt{emcee}$ MCMC sampler~\cite{emcee}. The best-fit values  are displayed in Fig.~\ref{fig:fig3},~\ref{fig:fig4}, and~\ref{fig:fig5} for the different cases presented in Sec.~\ref{case1},~\ref{case2}, and~\ref{case3} respectively. The diagonal entries in each figure represent the one dimensional marginalized posterior distributions of the  corresponding parameters used in each particular case and off-diagonal contour plots show the 68\%, 95\%, and 99\%  two-dimensional  marginalized credible intervals. We reported no violation of CDDR at $1\sigma$ for all the  cases. The values of $\eta_0$ studied in this work for different cases along with recent studies can be found in Table~\ref{tab:table1}.

\section{Conclusions}
\label{sec:conclusions}
 
In this work, we have proposed a new  analysis in order to search for any cosmological signatures of a possible breaking in Einstein equivalence principle in electromagnetic sector. As pointed by Ref.\cite{hees}, alternative theories described by the action(\ref{action1}) lead naturally to variations of the fine structure constant, departures of the cosmic distance duality relation and also modifications of the  CMB temperature evolution law. 

The cosmological data used here were 40 galaxy cluster gas mass fraction measurements and 31 $H(z)$ data. We explore the fact that the angular diameter distances of the clusters via X-ray observations depend on the fine structure constant and  of the cosmic distance duality relation validity. On the other hand, the angular diameter distance to each galaxy cluster was calculated by applying  Gaussian Processes Regression  to  $H(z)$ data.  A possible breaking of the  Einstein equivalence principle was parameterized by a function such as $\eta(z)=1+\eta_0z$, where $\eta_0=0$ corresponds to the  standard result of general relativity. The scenarios explored were: a flat universe with the depletion factor considered constant,  a flat universe with the depletion factor evolving over time and a non-flat universe with the depletion factor taken as a constant. For  all these cases,  we find that  $\eta_0=0$ within 1$\sigma$ c.l. and no departure of the standard framework was observed (see Table~\ref{tab:table1}). In the near future, eROSITA observations will provide significant gains over available X-ray surveys, with $\approx$ 100,000 galaxy clusters are expected to be detected in X-ray band~\cite{eROSITA}. Then, tighter limits can be performed on $\eta(z)$ by using the method proposed here in order to search for any cosmological signature of a possible breaking in Einstein equivalence principle in electromagnetic sector.

\section*{ACKNOWLEDGEMENT}
KB would like to thank the Department of Science and Technology, Government of India for providing the financial support under DST-INSPIRE Fellowship program. RFLH
acknowledges CNPq No.428755/2018-6 and 305930/2017-6. IECR thanks to CAPES.

\end{document}